\documentclass[aps,prb,twocolumn]{revtex4}
\usepackage[dvips]{epsfig,color}
\usepackage{graphicx}
\begin{document} 
\title{Anisotropic Hubbard model on a triangular lattice  ---
spin dynamics in $\rm Ho Mn O_3$}
\author{Saptarshi Ghosh}
\email{gsap@iitk.ac.in} 
\author{Avinash Singh}
\email{avinas@iitk.ac.in} 
\affiliation{Department of Physics, Indian Institute of Technology Kanpur - 208016}
\begin{abstract}
The recent neutron-scattering data for spin-wave dispersion in 
$\rm Ho Mn O_3$ are well described by an anisotropic Hubbard model on a 
triangular lattice with a planar (XY) spin anisotropy. 
Best fit indicates that magnetic excitations in $\rm Ho Mn O_3$ 
correspond to the strong-coupling limit $U/t > \sim 15$, 
with planar exchange energy $J=4t^2/U \simeq 2.5$meV and 
planar anisotropy $\Delta U \simeq 0.35$meV.
\end{abstract}
\pacs{75.50.Pp,75.30.Ds,75.30.Gw}  
\maketitle
There has been renewed interest in correlated electron systems on triangular lattices,
as evidenced by recent studies of antiferromagnetism, superconductivity 
and metal-insulator transition in the organic systems 
$\rm \kappa -(BEDT-TTF)_2 X$,\cite{review1,review2} 
the discovery of superconductivity in $\rm Na_x Co O_2 . y H_2 O$,\cite{watersup}
the observation of low-temperature insulating phases 
in some $\sqrt{3}$-adlayer structures such as K on Si[111],\cite{weitering} 
and quasi two-dimensional $120^0$ spin ordering and spin-wave excitations
in $\rm Rb Fe (MoO_4)_2$ (Refs. 5,6) 
and the multiferroic materials $\rm Y Mn O_3$ and $\rm Ho Mn O_3$.\cite{sato,holmium}

Recent neutron-scattering studies of the multiferroic material $\rm Ho Mn O_3$ 
have revealed a non-collinear $120^0$ antiferromagnetic (AF) ordering below
$T_{\rm N} \approx 72$ K of the $S=2$ Mn$^{3+}$ spins arranged in offset layers of
two-dimensional (2D) triangular lattice.\cite{holmium} 
Measurements of the spin wave dispersion were found to be well described by 
a nearest-neighbour Heisenberg AF with exchange energy $J = 2.44$ meV 
and a planar anisotropy $D=0.38$ meV at 20 K.
No discernible dispersion was observed in the out-of-plane direction,
indicating primarily 2D spin dynamics.

Recently spin-wave excitations in the $120^0$ AF state of the Hubbard model 
on a triangular lattice were studied within the random phase approximation (RPA)
in the full $U$ range.\cite{tri}
The spin wave energy in the large $U$ limit was shown to asymptotically approach 
the corresponding result for the Quantum Heisenberg antiferromagnet (QHAF), 
thus providing a continuous interpolation between weak and strong
coupling limits. However, competing interactions and frustration
were found to significantly modify the dispersion at finite $U$, 
resulting in vanishing spin stiffness at $U\approx 6$ 
and a magnetic instability at $U \approx 7$ corresponding to 
vanishing spin-wave energy at wave vector ${\bf q}_M=(\pi/3,\pi/\sqrt{3})$.
The sharp fall-off of $\omega_M$ near $U \approx 7$ provides a sensitive indicator
of finite-$U$ effects in the AF state. 
Indeed, recent high-resolution neutron-scattering studies of the spin-wave dispersion 
in the square-lattice S=1/2 AF $\rm La_2 Cu O_4$
have revealed noticeable spin-wave dispersion along the MBZ edge,\cite{spinwave} 
associated with finite-$U$ double-occupancy effects.\cite{spectrum} 

In this brief report we extend the spin-wave analysis to include planar spin anisotropy, 
and show that the neutron-scattering data for $\rm Ho Mn O_3$ 
are also well described by a Hubbard model on a triangular lattice,
thus providing a microscopic description of the most essential features 
of spin dynamics in the $120^0$ AF state of $\rm Ho Mn O_3$,
including the spin-wave dispersion and energy scale.
We examine the behaviour of spin-wave anisotropy gap with spin anisotropy
and also suggest a sensitive measure of finite-$U$, double occupancy effects. 

However, the Mn spin planar anisotropy is treated here only at a phenomenological level, 
equivalent to the effective anisotropy $DS_{iz}^2$ included 
in recent investigations using spin models.\cite{holmium,sato}
In a detailed study of single-ion anisotropy and crystal-field effects 
in the layered rare-earth cuprates $\rm R_2 CuO_4$ (R=Nd,Pr,Sm), 
the magnetic behaviour (including spin-reorientation transitions)
has been attributed to coupling of Cu with the rare-earth magnetic subsystem 
which exhibits a large single-ion anisotropy resulting in preferential ordering 
of rare-earth moments along specific lattice directions.\cite{sachi} 
It has been suggested that the anisotropy of Mn spins, 
its observed temperature dependence, 
and the reorientation transitions in $\rm Ho Mn O_3$ also originate from a similar 
anisotropic exchange coupling with the rare-earth Holmium,\cite{holmium}
resulting in magnetic behaviour as seen in layered rare-earth cuprates,
where frustrated interlayer coupling allows for weaker, higher-order interactions
to control the magnetic structure.
Especially relevant for non-collinear ordering,
the anisotropic exchange (Dzyaloshinski-Moriya) interaction 
${\bf D}.({\bf S}_i \times {\bf S}_j)$
originating from spin-orbit coupling has been suggested as responsible for the 
clamping of ferroelectric and antiferromagnetic order parameters in
$\rm Y Mn O_3$.\cite{hanamura}

Hund's rule coupling responsible for the $S=2$ spin state of Mn$^{+++}$ ions
and crystal-field splitting have also not been 
realistically incorporated here. 
However, these realistic details do not qualitatively affect 
the spin-rotation symmetry and spin dynamics, as discussed below.
Hund's rule coupling in the generalized Hubbard model considered here is maximal as
inter-orbital Coulomb interaction for parallel spins is dropped completely.
Including an inter-orbital density-density interaction $V_0$ and
an intra-atomic exchange interaction $F_0$ favouring parallel-spin alignment 
(Hund's rule coupling),
the more realistic orbital Hubbard model\cite{roth,held}
\begin{eqnarray}
H &=& -t \sum_{i,\delta,\gamma,\sigma} a_{i\gamma\sigma}^\dagger a_{i+\delta,\gamma\sigma} 
+ U\sum_{i\gamma} n_{i\gamma\uparrow} n_{i\gamma\downarrow} \nonumber \\
&+& 
\sum_{i,\gamma<\gamma',\sigma,\sigma'}(V_0 - \delta_{\sigma\sigma'}F_0 )
n_{i\gamma\sigma} n_{i\gamma'\sigma'} \nonumber \\
&+&
F_0 \sum_{i,\gamma<\gamma',\sigma\ne\sigma'}
a_{i\gamma\sigma} ^\dagger
a_{i\gamma'\sigma}
a_{i\gamma'\sigma'} ^\dagger
a_{i\gamma\sigma'}
\end{eqnarray}
remains spin-rotationally invariant under a global rotation of the fermion spin
${\bf S}_{i\gamma}=\Psi^\dagger _{i\gamma} \frac{\mbox{\boldmath $\sigma$}}{2} \Psi_{i\gamma}$
(where $\Psi^\dagger _{i\gamma} \equiv (a_{i\gamma\uparrow}^\dagger \; a_{i\gamma\downarrow}^\dagger)$ is the fermion field operator),
even if orbitals $\gamma$ are identified with the Mn orbitals $(t_{2g},e_g)$ 
resulting from crystal-field splitting of the atomic 3d orbitals.
As the intra-atomic exchange interaction $F_0$ is much larger than the 
spin excitation energy scale ($\sim \frac{t^2}{U+F_0}$, within a strong-coupling expansion), 
all fermion spins on a site are effectively coupled, 
yielding a composite quantum spin $S$ 
and a corresponding multiplication by factor $2S$
to obtain the effective spin-wave energy scale. 
Therefore, orbital multiplicity does not change the Goldstone-mode structure,
and the spin-dynamics energy scale in the orbital Hubbard model 
is essentially determined by $t$ and $U_{\rm eff} = U+F_0$.

As a simplest extension to phenomenologically include spin-space anisotropy, 
while retaining only the relevant energy scales $t$ and $U_{\rm eff}$,
we consider the generalized $\cal N$-orbital Hubbard model\cite{quantum} 
\begin{eqnarray}
H &=&  -t \sum_{i,\delta,\gamma,\sigma} 
a_{i\gamma\sigma} ^\dagger  a_{i+\delta,\gamma,\sigma} 
+ \frac{U_1}{\cal N} \sum_{i,\gamma,\gamma'} a_{i\gamma\uparrow}^\dagger 
a_{i\gamma\uparrow} a_{i\gamma'\downarrow}^\dagger a_{i\gamma'\downarrow} \nonumber \\
&+& \frac{U_2}{\cal N} \sum_{i,\gamma,\gamma'} a_{i\gamma\uparrow}^\dagger 
a_{i\gamma'\uparrow} a_{i\gamma'\downarrow}^\dagger a_{i\gamma\downarrow}
\end{eqnarray}
on a triangular lattice with nearest-neighbour (NN) hopping between sites $i$ 
and $i+\delta$. 
Here $\gamma,\gamma'$ refer to the (fictitious) degenerate $\cal N$ orbitals per site.
The factor $\frac{1}{\cal N}$ is included to render the energy density finite
in the ${\cal N} \rightarrow \infty$ limit. 
The two correlation terms involve density-density and exchange interactions 
with respect to the orbital indices. 
The Hartree-Fock (HF) approximation and Random Phase Approximation (RPA) are of O(1)
whereas quantum fluctuation effects appear at higher order within the 
inverse-degeneracy expansion and thus $1/\cal N$,
in analogy with $1/S$ for quantum spin systems, plays the role of $\hbar$. 

The key feature of spin-rotation symmetry of the generalized Hubbard model 
is highlighted by writing the two interaction terms as 
\begin{equation}
H_{\rm int} =  
- \frac{U_2}{\cal N} \sum_{i} {\bf S}_{i} . {\bf S}_{i} 
+ \frac{U_2 - U_1}{\cal N} \sum_{i} S_{iz} ^2 
\end{equation}
in terms of the total spin operator
\begin{equation}
{\bf S}_{i} =  \sum_\gamma \psi_{i\gamma}^\dagger 
\frac{\mbox{\boldmath $\sigma$}}{2} \psi_{i\gamma}
\equiv  \sum_\gamma \frac{\mbox{\boldmath $\sigma$}_{i\gamma}}{2}
\end{equation}
where $\psi_{i\gamma}^\dagger \equiv (a_{i\gamma\uparrow}^\dagger \; 
a_{i\gamma\downarrow}^\dagger )$. 
An Ising (uniaxial) anisotropy is obtained for $U_1 > U_2$,
a planar (XY) anisotropy for $U_2 > U_1$,
and full spin-rotation symmetry for $U_1 = U_2$.

As appropriate for $\rm Ho Mn O_3$,
we consider the case $U_2 > U_1$ corresponding to preferential ordering of spins
in the $x-y$ plane in spin space and an anisotropy gap for out-of-plane  
excitations.\cite{holmium}
Magnetic excitations were analyzed in terms of a 
Heisenberg model with a similar anisotropy term in Ref. [7].
At the HF level, the interaction term for orbital $\gamma$ then reduces to 
\begin{equation}
H_{\rm int} ^\gamma =  
- \sum_i \mbox{\boldmath $\sigma$}_{i\gamma} .{\bf \Delta}_i  
\end{equation}
where the self-consistently determined mean field 
${\bf \Delta}_i = U_2 \langle {\bf S}_{i\gamma'} \rangle_{\rm HF}$
lies in the $x-y$ plane in spin space.
The HF sublattice magnetization depends only on $U_2$ and is determined from the 
self-consistency condition $\langle {\bf S}_{\alpha} \rangle_{\rm HF}=
\sum_{{\bf k},l}\langle {\bf k},l| \frac{\mbox{\boldmath $\sigma$}}{2} 
|{\bf k},l \rangle_\alpha $ in terms of the HF states $|{\bf k},l \rangle$
on sublattice $\alpha$, exactly as for the isotropic case.\cite{tri}
In the strong coupling limit the sublattice magnetization 
$\langle {\bf S}_{\alpha} \rangle \approx 1/2$  
at the HF level, and is reduced by about 50\% (in the isotropic case) due to quantum fluctuations, as found in different calculations cited in Ref. [9].

We consider the $120^0$ ordered AF state on the triangular lattice,
and examine transverse spin fluctuations about the broken-symmetry state.
At the RPA level,  
the magnon propagator reduces to a sum of all bubble diagrams
where the interaction vertices involving  
$S_{ix}^2$ and $S_{iy}^2$ appear with interaction $U_2$,
whereas those involving $S_{iz}^2$ with interaction $U_1$.
Introducing a planar spin rotation,
so that spins are oriented along the $x'$ direction for all three sublattices, 
we obtain for the transverse spin-fluctuation propagator 
\begin{equation}
[\chi({\bf q},\omega)]_{\alpha\beta} ^{\mu\nu} = 
\frac{[\chi^0 ({\bf q},\omega)]}
{{\bf 1} - 2[U][\chi^0 ({\bf q},\omega)]} 
\end{equation}
in the $2\otimes 3$ spin-sublattice basis of the two transverse spin directions
$\mu,\nu=y',z'$ and the three sublattices $\alpha,\beta=A,B,C$.
The sublattice-diagonal interaction matrix 
\begin{equation}
[U] = \left [ \begin{array}{cc} U_2 {\bf 1} & {\bf 0} \\
{\bf 0} & U_1 {\bf 1} \end{array} \right ]
\end{equation}
in the $y',z'$ basis, and the bare particle-hole propagator 
\begin{eqnarray}
& & [\chi^0({\bf q},\omega)]_{\alpha \beta}^{\mu\nu}  \nonumber \\ 
&=& \frac{1}{4} \sum_{{\bf k},l,m} 
\left [ \frac{\langle \sigma_\mu \rangle_\alpha ^{-+}
\langle \sigma_\nu \rangle_\beta^{-+*}}
{E_{{\bf k-q},m}^+ - E_{{\bf k},l}^- + \omega} 
+ \frac{\langle \sigma_\mu \rangle_\alpha ^{+-}
\langle \sigma_\nu \rangle_\beta ^{+-*}}
{E_{{\bf k},m}^+ - E_{{\bf k-q},l}^- - \omega} \right ] \nonumber \\
\end{eqnarray}
involves integrating out the fermions in the broken-symmetry state.
In the particle-hole matrix elements 
\begin{equation}
\langle \sigma_\mu \rangle_\alpha ^{-+} \equiv
\langle {\bf k-q},m| \sigma_\mu |{\bf k},l\rangle_\alpha 
\end{equation}
of the rotated spins,
the spin orientation angles $\phi_\alpha$ in the fermion states 
$|{\bf k},l\rangle$ are transformed out.
The numerical evaluation of $[\chi^0({\bf q},\omega)]$ in terms of the HF-level
AF-state energies and amplitudes is exactly as for the isotropic case 
studied earlier.\cite{tri}

The spin-wave energies $\omega_{\bf q}$ are then obtained 
from the poles $1 - \lambda_{\bf q}(\omega_{\bf q}) = 0$ of Eq. (6)
in terms of the eigenvalues $\lambda_{\bf q}(\omega)$ 
of the matrix $2[U][\chi^0({\bf q},\omega)]$.
As $\omega_{\bf q}$ corresponds to spin 1/2,
it is scaled by the factor $2S$ for arbitrary spin $S$.\cite{magimp}
As expected for planar anisotropy, there is only one Goldstone mode corresponding
to planar rotation of spins, and the two out-of-plane modes become massive
with an anisotropy gap $\omega_{\rm gap}$.

\begin{figure}
\begin{center}
\vspace*{-00mm}
\hspace*{-5mm}
\psfig{figure=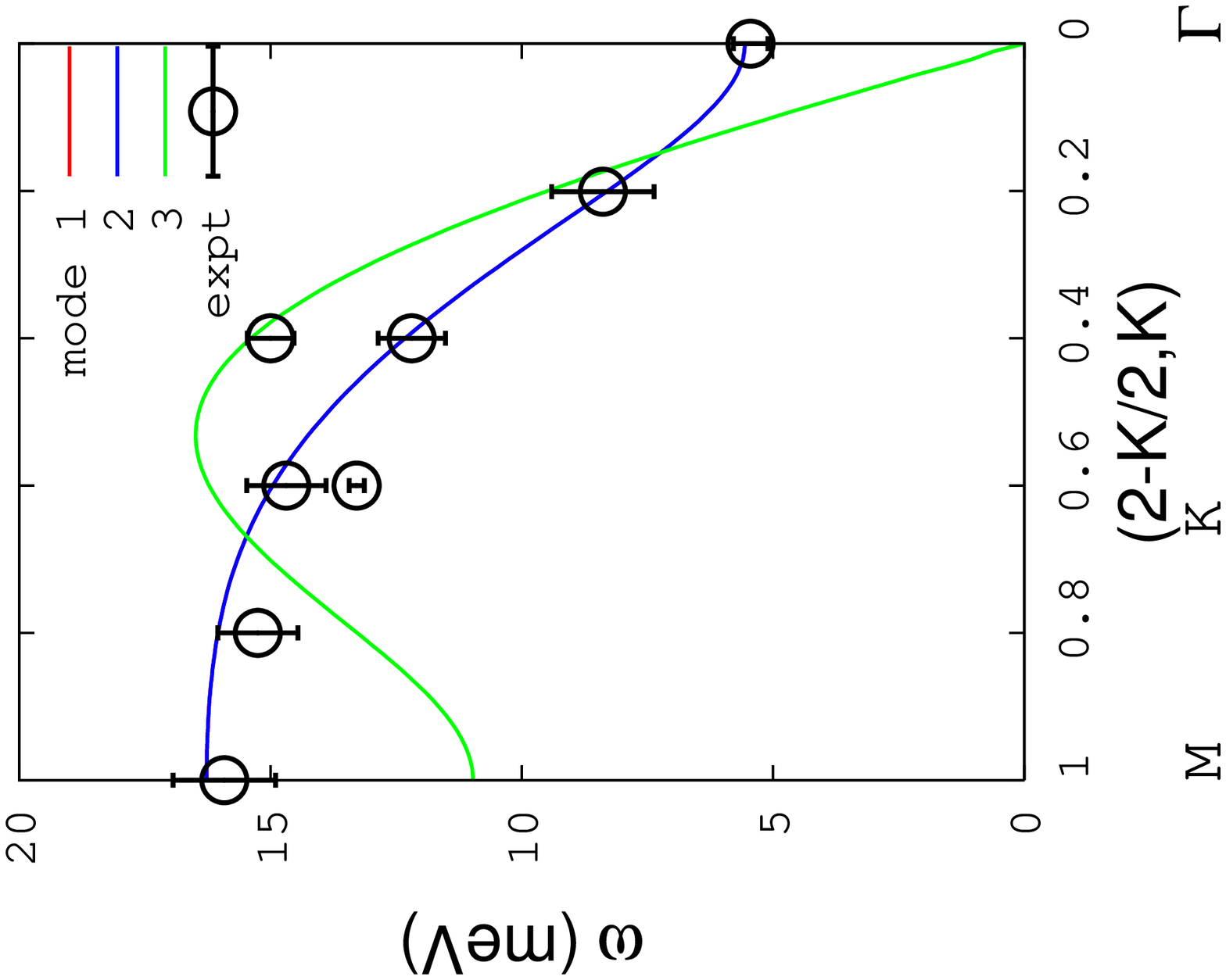,height=60mm,angle=-90}
\vspace{0mm}
\end{center}
\vspace*{-125mm}
\hspace*{-40mm}
\epsfig{figure=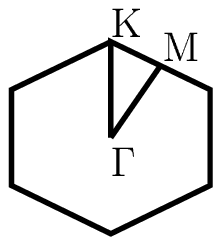,width=140mm}
\vspace{-75mm}
\end{figure}

\begin{figure}
\begin{center}
\vspace*{-10mm}
\hspace*{-5mm}
\psfig{figure=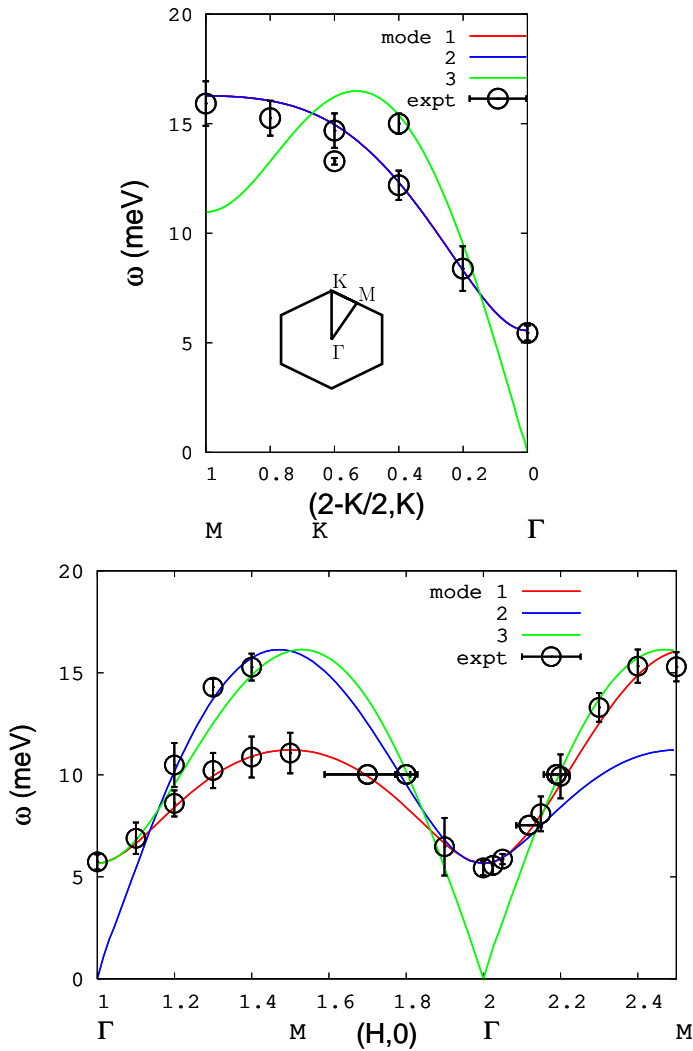,height=95mm,angle=-90}
\vspace{-00mm}
\end{center}
\caption{(color online) Spin-wave dispersion for the three modes 
calculated from Eq. (5) with $J=4t^2/U \approx 2.5$meV and $\Delta U \approx 0.35$meV,
along with neutron scattering data points for $\rm Ho Mn O_3$ at 20K from Ref. [7].}  
\end{figure}

Spin-wave dispersion $\omega_{\bf q}$ for the three modes is shown in Fig. 1 
along two symmetry directions in the magnetic Brillouin zone (MBZ),
along with the neutron-scattering data for $\rm Ho Mn O_3$ at 20 K from Ref. [7]. 
The anisotropic Hubbard model provides a remarkably good description 
of the spin dynamics.
We find that best fits with the magnetic excitations in $\rm Ho Mn O_3$
are only obtained in the strong-coupling limit $U/t > \sim 15$,
with a planar exchange energy $J =4t^2/U \simeq 2.5$meV and anisotropy
$U_2 - U_1 \equiv \Delta U \simeq  0.35$meV,
the individual values of $t$ and $U$ not being resolvable within 
experimental resolution. To estimate the order of magnitude of the ratio $U/t$, 
if we nominally take $U=1$eV, we obtain $t=25$meV and $U/t=40$.
Spin-wave dispersion calculated in the intermediate coupling regime
cannot be fitted with the neutron scattering data, 
indicating no evidence of finite-$U$ double occupancy effects, as discussed below.

\begin{figure}
\begin{center}
\vspace*{-05mm}
\hspace*{-5mm}
\psfig{figure=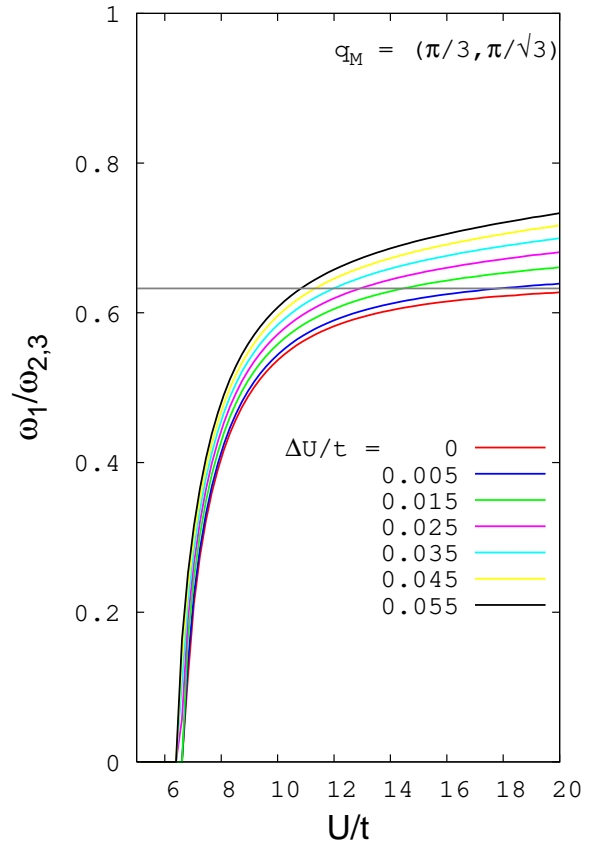,width=80mm}
\vspace{-00mm}
\end{center}
\caption{(color online) 
The ratio $\omega_1/ \omega_{2,3}$ at wave vector ${\bf q}_M$ 
provides a sensitive indicator of finite-$U$, double-occupancy effects, 
shown for different values of spin anisotropy $\Delta U$.} 
\end{figure}

\begin{figure}
\begin{center}
\vspace*{-5mm}
\hspace*{-5mm}
\psfig{figure=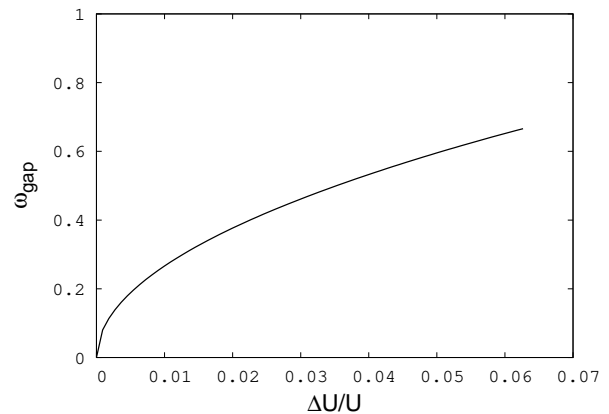,height=80mm,angle=-90}
\vspace{0mm}
\end{center}
\caption{Spin-wave anisotropy gap corresponding to out-of-plane fluctuations 
shows a $\sim (\Delta U/U)^{1/2}$ behaviour with spin anisotropy. Here $U/t=16$.}  
\end{figure}

For the square-lattice $S=1/2$ AF $\rm La_2 Cu O_4$, 
finite-$U$, double-occupancy effects associated with higher-order $(t^4/U^3$) 
spin couplings are manifested in noticeable 
spin-wave dispersion along the MBZ boundary.\cite{spinwave,spectrum}
Similarly, for the isotropic triangular-lattice AF,
the ratio $\omega_1/\omega_{2,3}$ of the non-degenerate and
degenerate spin-wave energies at wave vector ${\bf q}_M=(\pi/3,\pi/\sqrt{3})$
is a sensitive measure of the $U/t$ ratio,\cite{tri}
which asymptotically approaches $2/\sqrt{10} = 0.632$ in the strong-coupling 
$(U/t \rightarrow \infty)$ limit, decreases monotonically with $U/t$,
and eventually vanishes at $U/t \approx 7$.\cite{tri}
Variation of $\omega_1/ \omega_{2,3}$ with $U/t$ is shown in Fig. 2
for different values of anisotropy $\Delta U /t$.
Neutron-scattering data for $\rm Ho Mn O_3$ shows that 
$\omega_1/\omega_{2,3} \approx 11{\rm meV}/16{\rm meV} \approx 0.7$,
which yields a lower bound $(\sim 15)$ on the ratio $U/t$ from Fig. 2. 

The spin-wave anisotropy gap $\omega_{\rm gap}/t$,
corresponding to out-of-plane fluctuations with $q=0$, 
varies as $(\Delta U/U)^{1/2}$ with spin anisotropy (Fig. 3),
which translates to $\omega_{\rm gap} \propto \sqrt{J \Delta U}$,
the geometric mean of the two energy scales.

In conclusion, the anisotropic Hubbard model provides a simple extension to
phenomenologically include spin-space anisotropy and study spin excitations in 
the full range from weak to strong coupling.
Comparison of spin-wave dispersion with RPA calculations allows for a quantitative 
determination of the effective anisotropy, in addition to highlighting 
any finite-$U$, double occupancy effects as seen in $\rm La_2 Cu O_4$. 
The recent neutron-scattering data for spin-wave dispersion in 
$\rm Ho Mn O_3$ are found to be well described by an anisotropic Hubbard model on a 
triangular lattice with a planar (XY) spin anisotropy. 
We find that the twin constraints $\omega_{\rm gap}/\omega_{1,2} \approx 1/3$ 
and $\omega_1/\omega_{2,3} \approx 2/3$ in the neutron scattering data
cannot be satisfied in the intermediate coupling regime,
and best fit indicates that magnetic excitations in $\rm Ho Mn O_3$ 
correspond to the strong-coupling limit $U/t > \sim 15$, 
with $J \simeq 2.5$meV and $\Delta U \simeq 0.35$meV.
The ratio $\omega_1/\omega_{2,3}$ of the non-degenerate and
degenerate spin-wave energies at wave vector ${\bf q}_M=(\pi/3,\pi/\sqrt{3})$
is suggested as a sensitive measure of finite-$U$, double occupancy effects
in a triangular-lattice antiferromagnet.
Finally, in view of the formal resemblance with the 
orbital Hubbard model (1), our RPA calculation provides a step forward towards
investigating magnetic excitations using realistic models including Hund's rule 
coupling, crystal-field splitting etc.

\end{document}